*Anharmonic effects on lattice dynamics and thermal transport of two-dimensional InTe Monolayer*

*Hind Alqurashi\*[1,2], Abhiyan Pandit[3], and Bothina Hamad \*\*[3,4]*

[1] *Materials Science and Engineering graduate program, University of Arkansas, Fayetteville, AR 72701, USA*
[2] *Physics Department at the College of Science, Al Baha University, Al Baha 65527, Saudi Arabia*
[3] *Department of Physics University of Arkansas, Fayetteville, AR 72701, USA*
[4] *Physics Department, The University of Jordan, Amman-11942, Jordan*

## Abstract

The lattice thermal conductivity ($\kappa_l$) plays a key role in the performance of thermoelectric (TE) materials, where the lower values lead to higher figure of merit values. Two-dimensional (2D) group III-VI monolayers such as InTe are promising materials for TE energy generation owing to their low $\kappa_l$ that lead to high TE figure of merit values. In this work, we investigate the influence of the lattice anharmonicity on the lattice thermal conductivity of InTe monolayer. The thermodynamic parameters are calculated by using the self-consistent phonon (SCP) theory. The $\kappa_l$ value of the InTe monolayer is obtained to be 0.30 Wm$^{-1}$ K$^{-1}$ by using the standard Boltzmann transport equation (BTE) approach, while it is 3.58 Wm$^{-1}$ K$^{-1}$ by using SCP + BTE approach. These results confirm the importance of the anharmonic effects on the $\kappa_l$ value, where it was found to be significantly higher (91%) using the SCP + BTE approach than that obtained using the standard BTE approach.

*Corresponding author: \*h.hamidi@bu.edu.sa , \*\*bothinah@uark.edu*

# 1. Introduction

Lattice vibrations are considered to be one of the frontier research topics in materials science owing to their success in predicting the dynamic, thermodynamic and thermoelectric properties of crystalline solids [1,2]. The fundamentals of harmonic approximation (HA) have been very successful in the last few decades for predicting the thermodynamic properties. This method utilizes the second-derivative of the Born-Oppenheimer (BO) energy surface with the assumptions of relatively small atomic displacements [3]. The HA was sufficient for explaining the phonon dispersion curves, elastic properties, and lattice vibrations, but it fails to illustrate the anharmonic aspects such as the thermal expansion and the lattice thermal conductivity in crystalline solids [4].

The anharmonic effects are introduced by the cubic and the higher-order terms of the BO surface energy. These terms describe the phonon-phonon scattering, the relationship between phonon dispersion and finite temperature, the lifetimes of phonons [5,6] as well as the lattice thermal conductivity [5]. The anharmonic effects could be implemented using the density functional perturbation theory (DFPT) [4] and finite-displacement method [7], where cubic and higher order terms in finite-displacement method can be extracted from the force-displacement data. However, this method turns to be computationally expensive when the anharmonic order is increased due to the increase in range of adjacent atomic interactions [8,9]. In contrary, the anharmonic self-energies are considered as small perturbations in the DFTP method. This method is found to be valid only in the presence of small self-energies and it is unlikely to produce reliable results for variously anharmonic systems. Because of the imaginary frequencies of harmonic phonons, the DFPT fails in the high-temperature phase of ferroelectric materials [[10],[11],[12],[13]].

To surmount this constraint, the anharmonic effects can be treated using the nonperturbative ab initio molecular dynamic (AIMD) methods [14,15]. However, the drawback of these methods is due to the fact that most of them are based on Newton's equation of motion, which cannot account for the zero-point vibrations. Therefore, these methods cannot be used at low temperatures. One type of AIMD methods is the temperature-dependent effective potential (TDEP) method [16,17], which optimizes the effective harmonic force constants at finite temperatures. This method is efficient at high temperatures because it permits anharmonic terms to affect the phonon eigenvectors and the internal coordinate system. Although the TDEP works properly at finite temperatures, it fails in considering the zero-point vibrations at low temperatures. An alternative method of such AIMD methods is the self-consistent phonon (SCP) theory, which includes the anharmonic effects by incorporating the quantum effect of phonons in a nonperturbative approach [[18],[19],[20]]. The development of the effective implementation of SCP theory was achieved by employing force constants of higher order harmonics that are derived by the application of compressive sensing lattice dynamics (CSLD) approach [21].

The two-dimensional group III-VI monolayers have acquired a significant attention due to their promising thermoelectric properties at high temperatures [[22],[23],[24]]. For example, Mishra, *at.al.* predicted figure of merit (ZT) values of 1.01, 0.97, 0.90 for BSe, BS, BTe monolayers, respectively[25]. Moreover, a promising ZT value of 0.85 at 1100 K was predicted for GeTe hexagonal structure monolayer[26]. In addition, the transport properties of InTe two-dimensional monolayer were investigated using semiclassical Boltzmann Transport Equation (BTE) [27]. In this study, Çınar et al predicted a lattice thermal conductivity and a ZT value of 0.31WK$^{-1}$m$^{-1}$ and 3.94, respectively, at 1000K [27]. These values are promising for thermoelectric applications at high temperatures. However, lack of knowledge regarding the anharmonicity of lattice and its impact on the phonon dispersion curves and $\kappa_l$ values might lead to misleading information about the thermoelectric properties. The influence of lattice anharmonicity and finite temperature on thermal transport as well as the $\kappa_l$ of the InTe monolayer has not been addressed yet, which is the motivation of the current work. We utilized an effective computational method based on SCP theory [[28],[29],[30]] to compute the temperature-dependent phonon frequencies and $\kappa_l$ within a supercell by utilizing the interatomic force constants (IFCs).

The paper is arranged as follows: a brief information about the methodology is presented in section 2. Section 3 presents the computational methods. The results and discussions of the temperature dependence of the anharmonic phonon, lattice dynamics, and thermodynamic parameters are presented in section 4 and conclusion is presented in section 5.

## 2. Methodology

### 2.1 Potential energy expansion:

The Hamiltonian defines the dynamic of interacting nuclear system under the Born-Oppenheimer approximation, which is given as $H=T+U$, where $T$ and $U$ refer to kinetic and potential energies of the system, respectively. The potential energy ($U$) of the system can be expanded in Taylor series with regard to the atomic displacements ($u$) as follows [[5],[28],[31]]:

$$U = U_0 + U_2 + U_3 + U_4 + \cdots, \tag{1}$$

$$U_n = \frac{1}{n!}\Sigma_{\{\ell,k,\mu\}} \Phi_{\mu_1\ldots\mu_n}(\ell_1\kappa_1;\ldots;\ell_n\kappa_n) \times u_{\mu_1}(\ell_1\kappa_1)\cdots u_{\mu_n}(\ell_n\kappa_n), \tag{2}$$

Here, $U_n$, $u_\mu(\ell k)$, and $\Phi_{\mu_1\ldots\mu_n}(\ell_1\kappa_1;\ldots;\ell_n\kappa_n)$ are the nth-order contribution to the potential energy, the atomic displacement of the atom $k$ in the $\ell$th cell along the $\mu$ direction, and is the nth-order interatomic force constant (IFC), respectively. The linear term $U_1$ is neglected from Eq. (1)

owing to the zero atomic forces at equilibrium. Only the quadratic term $U_2$ is considered in the HA, whereas cubic, quartic, and higher-order terms are omitted. As a result, the Hamiltonian $U_0 = T + U_2$ can be expressed in terms of the harmonic phonon frequency ω. The dynamical matrix can be constructed to calculate the ω as [28]:

$$D_{\mu v}(\kappa\kappa'; \boldsymbol{q}) = \frac{1}{\sqrt{M_\kappa M_{\kappa'}}} \sum_{\ell'} \Phi_{\mu v}(\ell\kappa; \ell'\kappa') e^{i q \cdot r(\ell')}, \tag{3}$$

Here, $M_k$ refers to the mass of atom κ, $\Phi_{\mu v}(\ell\kappa; \ell'\kappa')$ represents the harmonic interatomic force constant, and $r(\ell')$ represents the primitive translation vector of the lattice. The Harmonic phonon frequency can be determined by diagonalizing the dynamical matrix, which is given as [28,32]:

$$\boldsymbol{D}(\boldsymbol{q})\boldsymbol{e}_{qj} = \omega_{qj}^2 \boldsymbol{e}_{qj}, \tag{4}$$

Here the $q$ and $j$ represent the wave vector and the phonon modes index, respectively, and the $\omega_{qj}$ and $e_{qj}$ are the phonon frequency and polarization vector of the phonon mode qj, respectively.

**2.2 Anharmonic self-energy and self-consistent theory**

The anharmonic contribution to the energy must be considered to explain the intrinsic scattering processes of phonon and the dependence of phonon frequency on temperature. The anharmonic terms can be treated as a perturbation $H'$ if they are small in comparison to the harmonic terms ($H_0 = T + U_0 + U_2$), which can be written as [[6],[28],[18]]:

$$H = H_0 + H' \approx H_0 + U_3 + U_4, \tag{5}$$

The higher-order terms (n>4) are eliminated because of their negligible contribution as compared to the cubic- and quartic-order terms. Nonetheless, nonperturbative treatment is required while drawing comparison between anharmonic and harmonic terms. The SCP theory defines a nonperturbative technique for dealing with anharmonic renormalization of phonon frequencies [[33],[28],[18]]. The Hamiltonian in Eq. (5) is modified to get the SCP equation as

$$H = \mathcal{H}_0 + (H_0 - \mathcal{H}_0 + U_3 + U_4) = \mathcal{H}_0 + \mathcal{H}', \tag{6}$$

where $\mathcal{H}_0$ is the effective harmonic Hamiltonian ( $\mathcal{H}_0 = \frac{1}{2}\sum_q \hbar\Omega_q \mathcal{A}_q \mathcal{A}_q^\dagger$) and $\Omega_q$ is phonon frequency after renormalization, $\mathcal{A}_q$ is the displacement operator, $\hbar$ refers to the Planck constant, and $q$ represents crystal momentum vector. The calculation of free energy system can be performed as the cumulative expansion of the term $\mathcal{H}'$. In addition, the principle of variation can be implemented using the first-order of SCP theory [6,30]. The SCP equation can therefore be obtained as follows:

$$\Omega_q^2 = \omega_q^2 + 2\Omega_q I_q, \tag{7}$$

$$I_q = \sum_{q_1} \frac{\hbar \Phi(q;-q;q_1;-q_1)}{4\Omega_q \Omega_{q_1}} \frac{[2n(\Omega_{q_1})+1]}{2}. \tag{8}$$

Here, $\omega_q$, and $\Phi(q;-q;q_1;-q_1)$ represent to the harmonic phonon frequency and the fourth-order IFC, and $\tilde{n}_{q_1} = \frac{1}{e^{\frac{\hbar\Omega_{q_1}}{k_B T}}-1}$ represents Bose–Einstein (BE) distribution function, $k_B$ represents Boltzmann constant and $T$ represents absolute temperature. The anharmonic phonon frequencies $\Omega_q$ can be found by using equations. (7) and (8).

## 2.3 Lattice thermal conductivity

The lattice thermal conductivity, $\kappa_l$, plays a significant role in improving the figure of merit $ZT$ in thermoelectric materials. It can be calculated using the Boltzmann transport equation within the relaxation time approximation (RTA) [6,34] as follows:

$$\kappa_l^{BTE} = \frac{\hbar^2}{N_q V k_B T^2} \sum_q \omega_q^2 v_q \otimes v_q n_q (n_q + 1) \tau_q, \qquad (9)$$

Here, $v_q = \frac{\partial \omega_q}{\partial q}$, is the group velocity, $n_q$ is distribution function of BE, $\tau_q$ is lifetime of quasi-particle accompanying phonon frequency, and $V$ is volume of the unit cell. The anharmonic effects are treated perturbatively. However, within the SCP theory the $\kappa_l$ can be written as[6,28]:

$$\tilde{\kappa}_l^{SCP+BTE} = \frac{\hbar^2}{N_q V k_B T^2} \sum_q \Omega_q^2 \bar{v}_q \otimes \tilde{v}_q \tilde{n}_q (\tilde{n}_q + 1) \tilde{\tau}_q, \qquad (10)$$

$$v_q = \frac{\partial \Omega_q}{\partial q}, \qquad (11)$$

$$\tilde{n}_q = n_q(\Omega_q), \qquad (12)$$

$v_q$ and $\tilde{n}_q$ are the terms illustrating the renormalized phonon frequency $\Omega_q$, and $\tilde{\tau}_q$ represents the renormalized lifetime illustrating the three-phonon scattering processes.

## 3. Computational details

The calculations are performed using the density functional theory (DFT) as implemented in VASP package [35]. The Perdew-Burke-Ernzerhof (PBE) approximation with generalized gradient approximation (GGA) was used to treat the exchange correlation functional [36]. In these computations, the plane waves were expanded up to cut-off energy of 520 eV with a total energy tolerance of $10^{-6}$ eV. A k-point mesh of $20 \times 20 \times 1$ is used with the van der Waals interactions included [37]. Along the *z*-direction, a 25 Å vacuum is included to avoid interactions between adjacent layers. These optimized parameters are then utilized for the electronic band structure calculations.

The IFCs were extracted using a 5 × 5 × 1 supercell of InTe monolayer with 100 atoms. The finite-displacement method was utilized to extract the harmonic IFCs. Each atom is displaced from its equilibrium location by 0.01Å, which considers all potential nearest-neighbor interactions. The calculations were performed up to the nineth nearest neighboring atomic interaction for the cubic IFCs. The cubic interaction force constants were obtained by employing the ordinary least squares (OLS) fitting technique with the harmonic interaction force constants, as applied in the ALAMODE package[[28],[6],[38]], after the atomic forces on each of the displaced configurations were computed. The CSLD method [21] is used to extract the fourth-order IFCs. This method is based on machine learning programs that have been addressed and applied in references [6,28]. A 5× 5 ×1 supercell of InTe monolayer was used to perform the AIMD calculations at room temperature for 14000 MD steps with a 1.5 fs time step. fifty equally spaced atomic structures were obtained from the trajectory of AIMD models. Afterwards, all of the atoms in each of the atomic structures were randomly displaced by 0.1 Å. The atomic forces for these structures were then estimated through density functional theory calculations. These computed atomic forces were then utilized to obtain the fourth-order of interaction force constants depending on the least absolute shrinkage and selection operator (LASSO) method [5] as follows:

$$\widetilde{\Phi} = \arg\ min_{\Phi} \|A\Phi - \mathcal{F}_{\text{DFT}}\|_2^2 + \alpha \parallel \mathbf{\Phi} \parallel_1, \tag{11}$$

where, $\Phi = [\Phi_1, \Phi_2, ..., \Phi_M]^{\text{T}}$ is a vector composed of $M$ linearly independent IFCs, $\mathcal{F}_{\text{DFT}}$ and $A$ refer to the vector of atomic forces and the matrix of atomic displacements, respectively. The ideal value of the hyperparameter $\alpha$ is obtained by using the cross-validation (CV) method. The harmonic term of interaction force constants is fixed to the values determined by the OLS approach, where the anharmonic term of interaction force constants is optimized by the LASSO regression step. After solving Eqs. (7) and (8), the predicted fourth-order IFCs were utilized to derive the anharmonic phonon frequency ($\Omega_q$). Thermodynamic parameters are calculated with $\Omega_q$ in this case. The specific heat capacity (Cv) is computed using the following formula:

$$C_V = \frac{k_B}{N_q} \sum_{q,j} \left(\frac{\hbar\Omega_{q_j}}{2k_B T}\right)^2 \operatorname{cosec} h^2 \left(\frac{\hbar\Omega_{q_j}}{2k_B T}\right), \tag{12}$$

Here, $N_q$ is the number of $q$-points.

## 4. Results and discussions

### 4.1. Structural and electronic properties

The two-dimensional InTe monolayer possesses a hexagonal crystal configuration with a space group $P\bar{6}m2$ (*no.* 187) containing four atoms in the primitive cell, as shown in Figure 1. The optimized lattice constant for this monolayer is found to be 4.39 Å, which is in agreement with previous ab initio investigations [39,40], see Table 1. The vertical distance between telluride atoms ($d_{Te\text{-}Te}$), the distance between indium and telluride atoms ($d_{In-Te}$), and the vertical distance between indium atoms ($d_{In-In}$) of InTe monolayer are also presented in Table1. These results are comparable to other previous calculations [39,40]. Figure 2 presents the electronic band structure along the high symmetry k-path and the Brillouin zone used in these calculations. Figure 2 (a) shows that InTe monolayer has an indirect band gap of 1.32 eV. The valence band maximum is found to be between $K$ and $\Gamma$ high-symmetry points, whereas the conduction band minimum is located at the $\Gamma$ high-symmetry point. The calculated band gap value (1.32 eV) is found to be in agreement with previous theoretical investigations of 1.34eV [27], 1.32eV [40] and 1.29eV [41].

### 4.2. Anharmonic force constants using LASSO

The SCP calculations were performed using the harmonic and fourth-order IFCs. The fourth-order IFCs were determined using the LASSO technique. For these calculations, 60 displacement-force data sets were set employing AIMD. The four-fold cross-validation (CV) technique was used to examine the predictive accuracy of LASSO regression [42]. Figure 3(a) presents the relative error of the atomic forces as the hyperparameter ($\alpha$) function. From this figure, one can see that the difference between the training and CV errors is negligible at large $\alpha$. In addition, the CV error decreases by decreasing $\alpha$ and reaches to its minimum value of $1.31 \times 10^{-6}$. The value of $\alpha$ is represented by the dotted vertical line in the figure, which is selected as an estimate of the fourth-order IFCs to provide the optimal accuracy for the data sets. The number of non-zero fourth-order IFCs is presented in Figure. 3(b). With the optimal $\alpha$ value, the total number of non-zero fourth-order IFCs is found to be 4104, which is about 84% of the total number of fourth-order IFCs.

**4.3. SCF at finite temperature**

This subsection presents the phonon dispersion relation and total density of states (TDOS) of InTe monolayer with the Harmonic and SCF methods, see Figure 4. The calculations of the anharmonic frequencies were obtained using a $5 \times 5 \times 1$ $q$-point grid. Through the SCP calculations, the dielectric constants and Born effective charges were used to account for the non-analytic correction. The unit cell of InTe monolayer has four atoms, which gives twelve phonon modes with three acoustic modes at lower frequencies, and nine optical modes at the higher frequencies, see Figure 4(a). The three acoustic modes consist of two in-plane (longitudinal (LA), and transverse (TA)), and one out-of-plane (flexural acoustic (FA)) modes. Near the $\Gamma$ point, the LA and TA modes are linear, but the FA mode is flexural, which is comparable to other 2D monolayer materials such as $ZrS_2$ [5], graphene [43] InY (Y= Se, Sn, and Te) [39] and phosphorene [44]. This flexural characteristic is common in two-dimensional structures [45]. These HA findings are in a good agreement with previous theoretical results [39]. The inclusion of the fourth order

IFCs in the SCP technique leads to an increase in acoustic and optical modes of InTe monolayer. The HA phonon frequency of the low-energy optical mode at $M$ point is 48.54 cm$^{-1}$, which increases to 50.23 cm$^{-1}$ when the fourth order IFCs is included in the SCP method at 0 K. The phonon frequency reaches 50.71 cm$^{-1}$ at room temperature as shown in Table 2 due to the temperature-dependent factor $I_q$ that is included in the SCP calculations, see equations (7) and (8). Figure 4(b) presents the phonon TDOS as a function of frequency. In general, the contribution of the individual atomic masses to the lattice vibrations in a compound is crucial. The heavier atom contributes more to the phonon frequency of low-energy acoustic mode, and the lighter atom contributes more to the phonon frequency of the high-energy optical mode [46]. As seen in Figure 4 (b), there are two peaks one in the phonon frequency of low-energy acoustic mode and another in the phonon frequency of high-energy optical mode, which belong to the heavier atom of Te and the lighter atom of In, respectively.

### 4.4. Thermodynamic parameters

This subsection presents the specific heat capacity ($C_v$), mean-square displacement (*MSD*), total vibrational free energy ($E_{total}$), phonon mode-dependent Grüneisen parameter ($\gamma_{qj}$), cumulative phonon group velocities ($v_g$), and phonon lifetime ($\tau$) of InTe monolayer. The $C_v$ is a significant thermodynamic variable that contributes directly to the $\kappa_l$ ($\kappa_l \propto C_v$). Figure 5(a) depicts the change in $C_v$ with temperature using the SCP method. This figure shows low $C_v$ values at low temperatures, which indicates a lower contribution to the $\kappa_l$. A similar behavior was reported for BaZrS$_3$ chalcogenide perovskite [47]. The *MSD* of atoms in a system is an essential parameter that indicates their divergence from the equilibrium position. The average mean-square displacement tensor of atom $k$ is calculated as follows [5]:

$$\langle u_\mu^2(k)\rangle = \frac{\hbar}{M_k N_q}\sum_{q,j}\frac{1}{\Omega_{qj}}|e_\mu(k;q_j)|^2\left(\tilde{n}_{q_j}+\frac{1}{2}\right), \tag{13}$$

Here, $M_k$, and $e_\mu(k;q_j)$ refer to the atomic mass of atom $k$, and the corresponding atomic polarization. Figure 5(b) shows that the *MSD* value increases with temperature, which can be attributed to the enhanced heating impact at higher temperatures. This increase in *MSD* of thermal vibrations leads to a reduction in the thermal transport. The *MSD* value of In atom is found to be higher than that of Te atom due to the inverse proportionately between the mass and atomic displacement as shown in equation 13. This behavior can be more prominent when the temperature increases. In Figure. 5 (c), the effect of the SCP correction on the free energy ($E_{Total} = E_{QHA} + E_{SCP}$) is found to be insignificant at low temperatures (below 200 K), while it becomes more pronounced at higher temperatures. For example, the SCP correction energy, decreases from -2.2×10$^{-4}$ eV at room temperature K to -1.9×10$^{-3}$ eV at 800K. This reduction of the total vibrational free energy upon applying the SCP correction means that the system is more stable when the quartic anharmonicity is considered. This observation emphasizes the significance of anharmonic frequency renormalization in terms of thermal properties. Another significant parameter that evaluates the anharmonicity of the structure is the $\gamma_{qj}$. Using cubic IFCs, this parameter is calculated as [5]:

$$\gamma_{qj} = -\frac{\partial(\log\omega_{qj})}{\partial(\log V)}, \tag{14}$$

where V refers to the volume. From equation 14, a positive value of $\gamma_{qj}$ means that the frequency of the phonon mode decreases as a function of volume. Figure 6 (a) shows the computed $\gamma_{qj}$ of InTe monolayer as a function of phonon frequency. For the harmonic lattice dynamics, $\gamma_{qj}$ has negative and positive values in the low-energy region (acoustic phonon modes), while it has a positive value in the high-energy region (optical phonon modes). This trend of $\gamma_{qj}$ in the case of InTe monolayer is similar to those of other systems such as $Cu_2O$ [48]. For the SCP, the value of $\gamma_{qj}$ has a negative value in the low-frequency region and a positive value in the high-frequency region. This implies that phonon anharmonicity is greater in the case of acoustic phonon modes. According to the continuum theory [49,48], this nature has a strong influence on the $\kappa_l$ via phonon lifetime ($\tau$) as $\tau_{qj}^{-1} \propto \gamma_{qj}^2$ [5]. Figure 6 (b) shows the $v_g$ of InTe monolayer as a function of phonon frequency. The $v_g$ values in the low-energy region are found to be greater than those of the high-energy region. As a result of the relation $\kappa_l \propto v_g$, the contribution of low-energy region to the $\kappa_l$ should be greater than that of high-energy region. Another important variable is the phonon lifetime ($\tau$), which is proportional to the $\kappa_l$ according to $\kappa_l \propto \tau$ [6,34]. Figure 6 (c) presents $\tau$ of InTe monolayer as a function of frequency at room temperature. From this figure, one can notice that the acoustic modes exhibit a longer phonon lifetime than that of the optical modes owing to the ratio of low phonon–phonon scattering. This indicates that the acoustic modes have a large impact in transporting most of the heat in InTe monolayer. This also plays a substantial contribution to the $\kappa_l$. The average value of $\tau$ is found to be 13.49 ps using SCP approach, while it is found to be 1.76 ps by using the harmonic approach. This can be attributed to the three-phonon scattering processes contained in the SCP lattice dynamics [5].

### 4.5. Lattice thermal conductivity

Figure 7 presents the lattice thermal conductivity ($\kappa_l$) spectrum and cumulative $\kappa_l$ with phonon frequency of InTe monolayer at room temperature using the BTE and SCP + BTE approaches. The spectrum of thermal conductivity in Figure 7 (a) shows that there are two peaks of phonons below 70 cm$^{-1}$. These two peaks are located in the acoustic and low energy of optical modes. Figure 7(b) presents the cumulative $\kappa_l$ as a function of phonon frequency. This figure shows that the contributions of phonons to the total thermal conductivity are about 91% and 97% using BTE and SCP + BTE approaches, respectively with frequencies lower than 70 cm$^{-1}$. However, the effect of higher frequency phonons (greater than 70 cm$^{-1}$) is almost negligible. This indicates that the acoustic and low energy optical modes of InTe monolayer play a significant role in the $\kappa_l$ value. This finding is in agreement with similar structures such as InSe, GaSe and GaS monolayers[50] as well as silicon and germanium two-dimensional systems [51].

Figure 8 depicts the contribution of several phonon branches to the $\kappa_l$ values. The acoustic phonon modes (1 to 3 modes) and low energy optical modes (4 and 5 modes) have more contributions to the $\kappa_l$ value than those of higher optical modes from 6 to 12 in both BTE and SCP + BTE approaches. Figure 9 presents the temperature dependence of $\kappa_l$ value estimated by the standard BTE and SCP + BTE techniques of InTe monolayer. The $\kappa_l$ computations were performed using a 100 ×100×1 $q$- points grid. The $\kappa_l$ value is found to decrease as a function of temperature. in conformity with the conventional relationship ( $\kappa_l \propto \frac{1}{T}$). In the case of the BTE method, the average $\kappa_l$ value of InTe monolayer at room temperature was found to be 0.30 Wm$^{-1}$ K$^{-1}$. This results is in agreement with a previous theoretical value of 0.31 Wm$^{-1}$ K$^{-1}$ using the same BTE approach [27]. The computed $\kappa_l$ values using the SCP + BTE approach are found to be significantly higher (91%) than those found using the standard BTE approach. At room temperature, the value of average $\kappa_l$ is found to be 3.58 Wm$^{-1}$ K$^{-1}$ by using the SCP + BTE approach, which is larger than that of the experimental value of InSe (about 1.6 Wm$^{-1}$ K$^{-1}$)[52]. These difference in the $\kappa_l$ values come from the difference in their phonon lifetime values. The higher $\tau$ values lead to the higher $\kappa_l$ values, see equations (9) and (10). The SCP+BTE approach with the higher $\tau$ values also show higher $\kappa_l$ values. These findings are consistent with previous anharmonic lattice dynamics calculations of SrTiO$_3$ [28], Ba$_8$Ga$_{16}$Ge$_{30}$[29], Ba$_8$Si$_{46}$ [53] and SCF$_3$ [54] using the SCP + BTE technique, where the computed findings were in agreement with the experimental results.

The SCP + BTE technique is considered to be more reliable than the standard BTE approach to estimate the $\kappa_l$ value, which leads to a more accurate prediction of the thermoelectric figure of merit and power efficiency. This can be related to the fact that the standard BTE method ignores the temperature reliance of phonon frequency and eigenvector. Therefore, it is unable to estimate the $\kappa_l$ value of high-temperature phase transitions due to the imaginary modes within the HA.

## 5. Conclusion

The InTe lattice dynamical properties are calculated within the SCP +BTE theory. In these calculations, the CSLD technique is used to find the higher-order harmonic IFCs. The Nonperturbative SCP approach is used to obtain the temperature-dependent phonon frequencies renormalized with the quartic anharmonicity. The phonon dispersion curves using the SCP approach are slightly higher than those of the Harmonic approach. In addition, the cumulative $\kappa_l$ values demonstrated that the most contribution of heat transfer is due to acoustic modes and low energy optical modes. Using the SCP + BTE approach, the $\kappa_l$ values are found to be higher than those obtained using the standard BTE approach. The SCP + BTE approach is believed to be more valid and accurate than the standard BTE approach in predicting the $\kappa_l$ value. This is due to the fact that the standard BTE approach eliminates the temperature reliance of phonon frequency and eigenvector. These results provide important insights into the effect of phonon anharmonicity on the lattice dynamics, thermodynamic properties and $\kappa_l$ value of the two- dimensional InTe monolayer, which play a key role in determining the thermoelectric properties.

**Acknowledgment**

Hind Alqurashi was financially supported by Al-Baha University and the Saudi Arabian Cultural Mission. The calculations were performed on the high-performance computing of the University of Arkansas.

**Conflict of interest**

The authors declare that they have no conflict of interest.

**TABLE 1.** The lattice constants (a), the distance between telluride atoms ($d_{Te\text{-}Te}$), the distance between Indium and telluride atoms ($d_{In-Te}$), and the distance between Indium atoms ($d_{In-In}$) for the InTe monolayer.

| InTe monolayer | a(Å) | $d_{Te\text{-}Te}$ (Å) | $d_{In\text{-}Te}$ (Å) | $d_{In\text{-}In}$(Å) |
|---|---|---|---|---|
| This work | 4.39 | 5.65 | 2.90 | 2.82 |
| Previous calculations | 4.38[39][40] | 5.57[39], 5.59[40] | 2.88[39], 2.88[40] | 2.81[39], 2.82[40] |

**TABLE 2.** Phonon frequencies (cm$^{-1}$) of the low-energy optical modes for InTe monolayer investigated utilizing different methods.

| Phonon branches | HA | SCP (at 0 K) | SCP (at 300 K) |
| --- | --- | --- | --- |
| 4 | 48.54 | 50.23 | 50.71 |
| 6 | 91.34 | 92.15 | 92.22 |

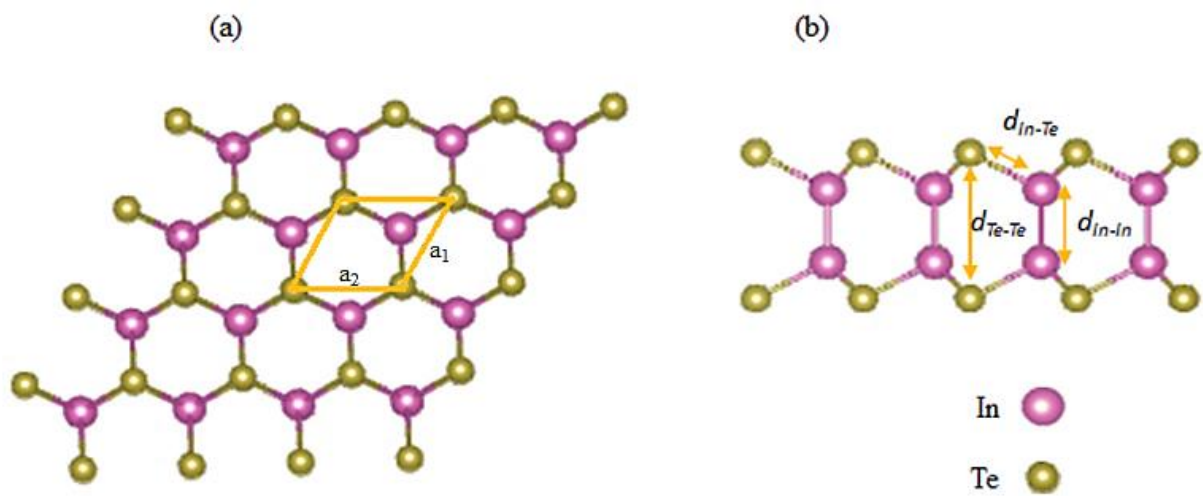

**Figure 1.** Crystal structure of InTe monolayer: (a) top view, the primitive unit cell is indicated in yellow, where $a_1=a_2$. (b) side view.

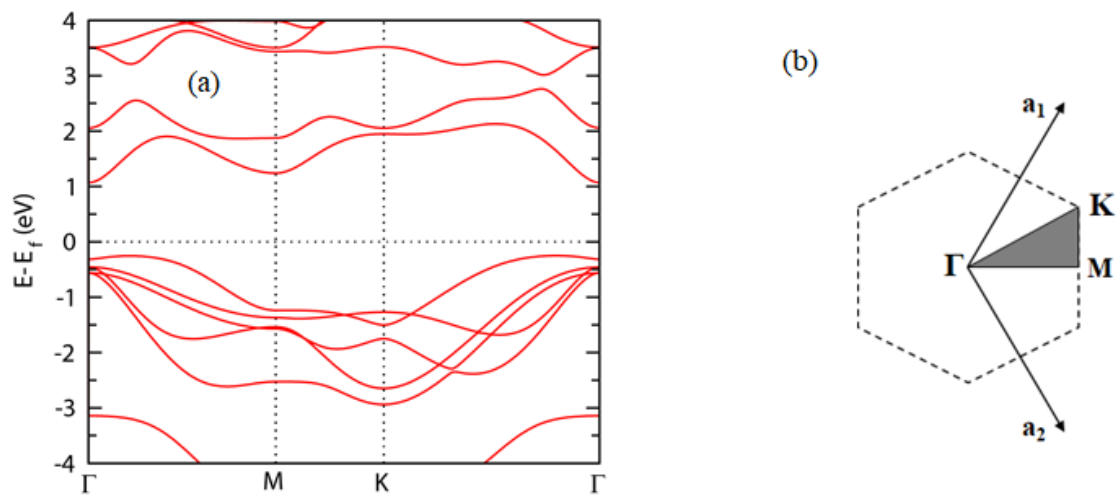

**Figure 2.** (a)Electronic band structure for InTe monolayer along the high-symmetry points in the first Brillouin zone (Γ–M–K–Γ) (b) The Brillouin zone with labeled high-symmetry points.

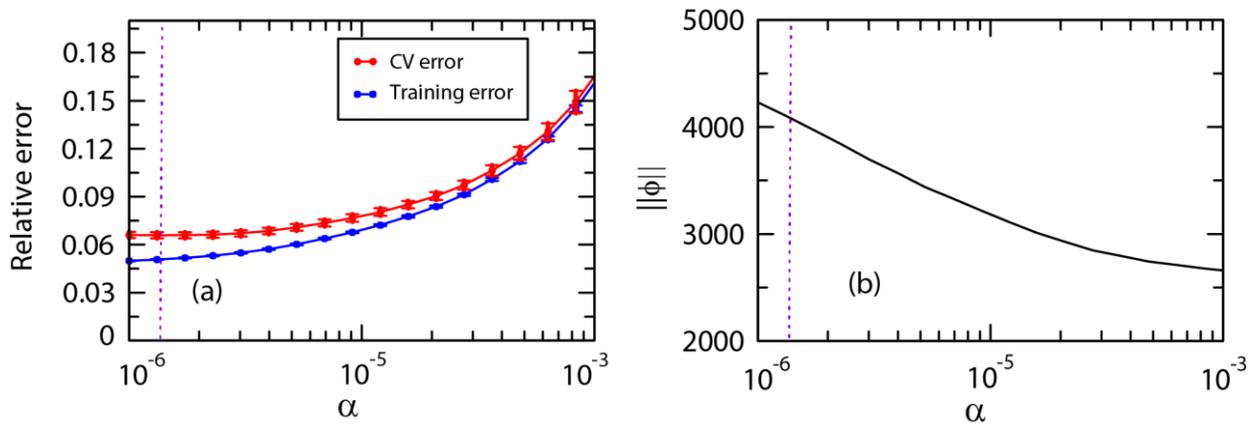

**Figure 3.** (a) Relative errors in the atomic forces and (b) the number of non-zero quartic IFCs with the hyperparameter ($\alpha$). The dotted vertical line refers to the value of hyperparameter ($\alpha$).

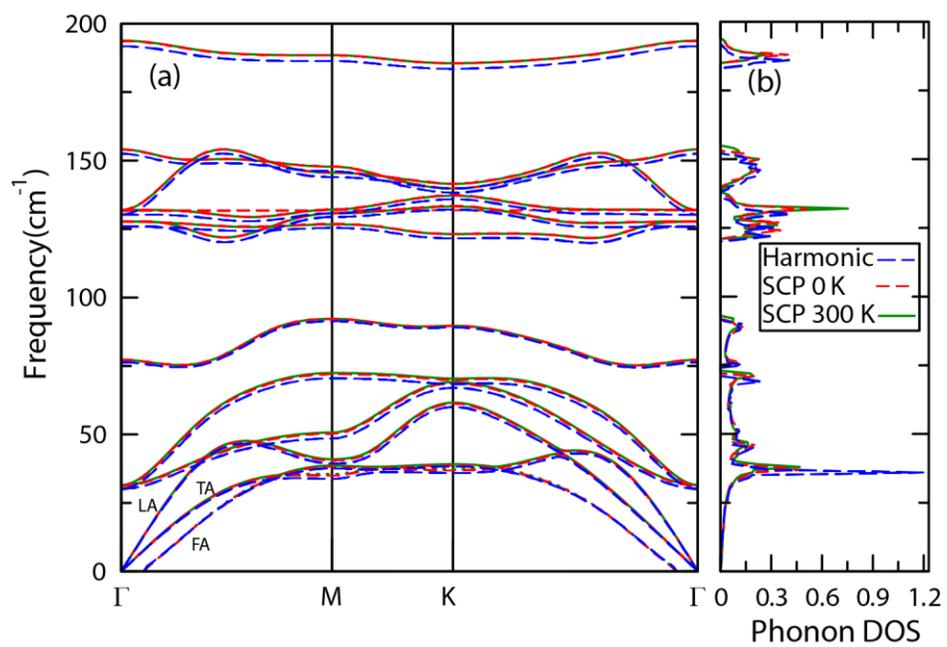

**Figure 4.** (a) Phonon dispersion relation and (b) the phonon DOS for the InTe monolayer obtained with the harmonic and SCP lattice dynamics.

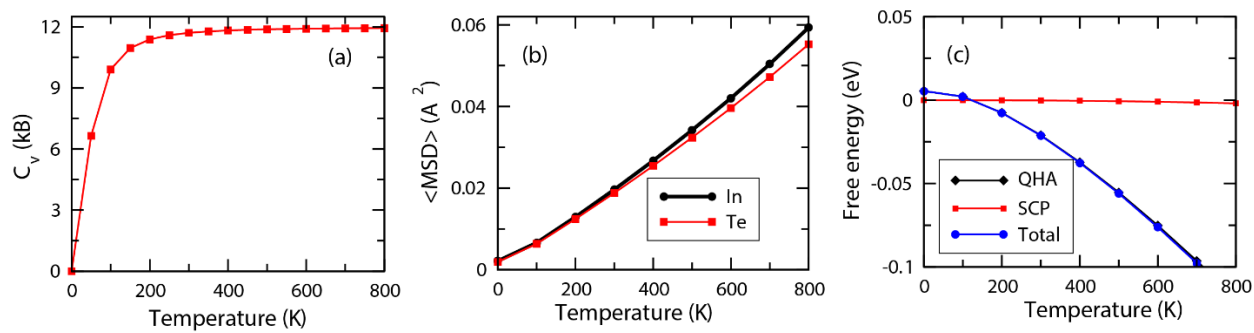

**Figure 5.** (a) The specific heat capacity ($C_v$) and (b) the *MSD* for the In and Te atoms, and (c) free energies within the QHA and SCP correction for the InTe monolayer with the temperature.

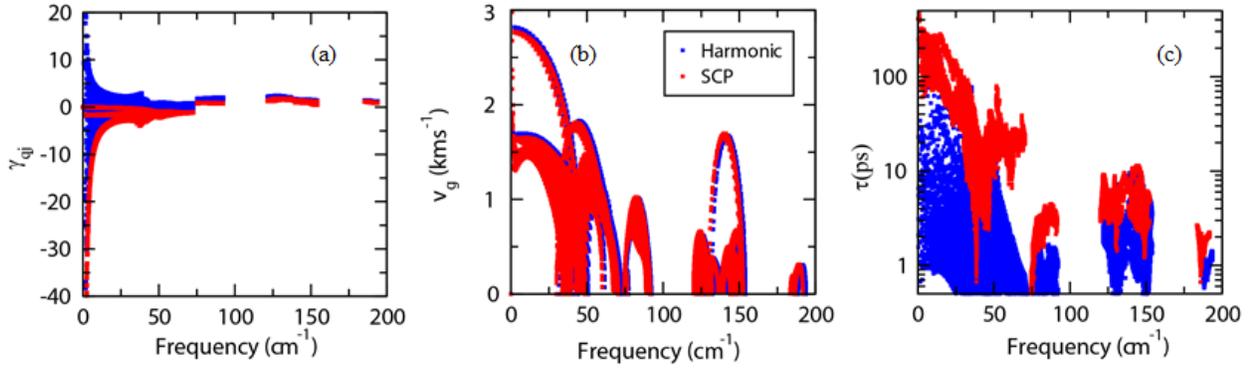

**Figure 6.** (a) Grüneisen parameter, (b) the cumulative phonon group velocity, and (c) the phonon lifetime of InTe monolayer with phonon frequency achieved with the harmonic and SCP lattice dynamics.

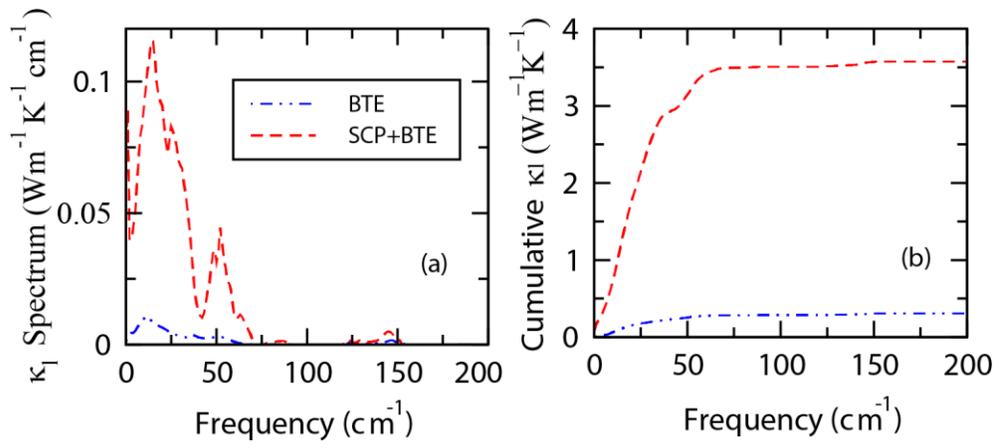

**Figure 7.** (a) lattice thermal conductivity ($\kappa_l$) spectrum and (b) cumulative $\kappa_l$ as with the phonon frequency of InTe monolayer at room temperature K achieved with the BTE and SCP + BTE approaches.

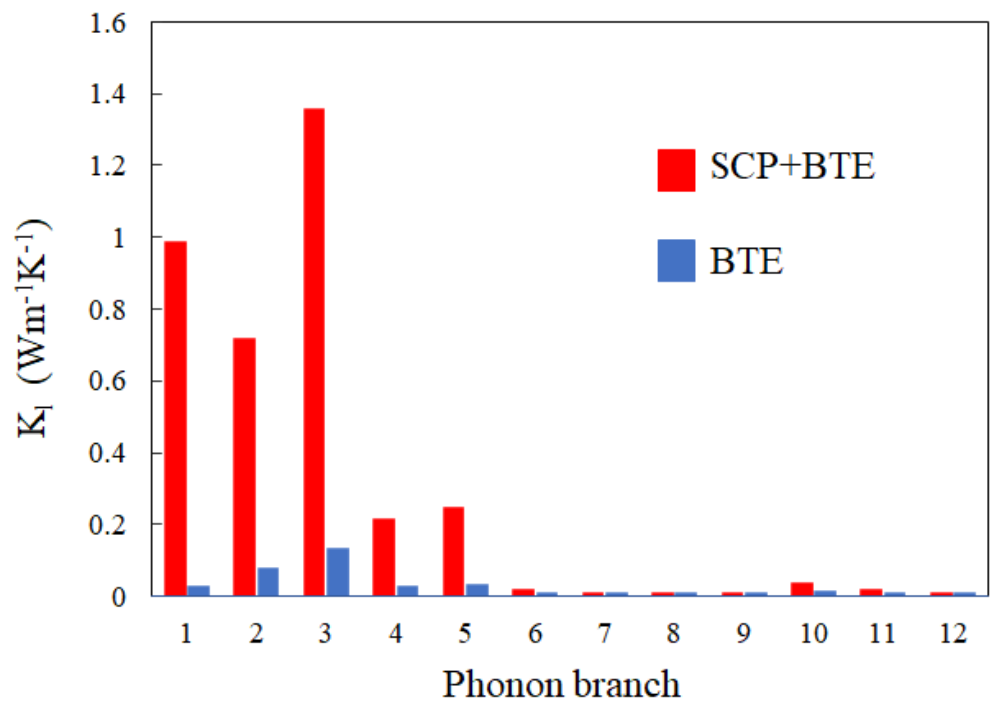

**Figure 8.** The total lattice thermal conductivity ($\kappa_l$), contribution of the phonon branches to $\kappa_l$ for InTe monolayer at room temperature K achieved with the BTE and SCP + BTE approaches.

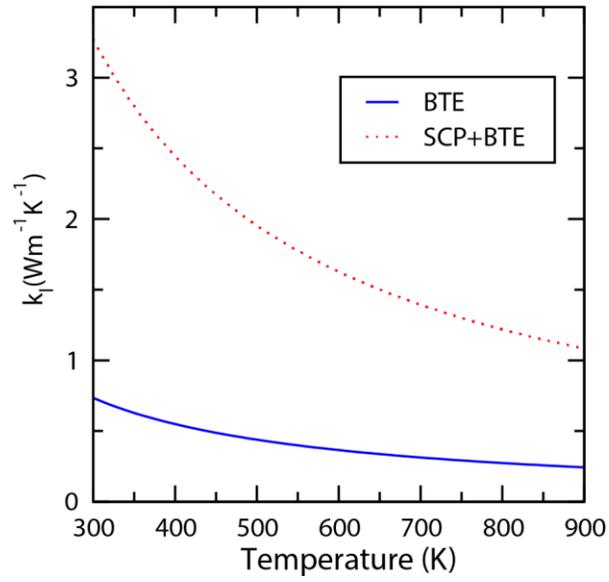

**Figure 9.** lattice thermal conductivity ($\kappa_l$) with temperature for the InTe monolayer achieved with the BTE and SCP + BTE approaches.